\documentclass[journal=jacsat,manuscript=article]{achemso}

\usepackage[version=3]{mhchem} % Formula subscripts using \ce{}
\usepackage{graphicx}
\usepackage{xcolor}
\usepackage{ulem}
\author{Laxmipriya Nanda}
\affiliation{School of Physical Sciences, National Institute of Science Education and Research, An
OCC of Homi Bhabha National Institute, Jatni-752050, Odisha, India}
\author{Bidyadhar Das}
\affiliation{School of Physical Sciences, National Institute of Science Education and Research, An
OCC of Homi Bhabha National Institute, Jatni-752050, Odisha, India}
\author{Subhashree Sahoo}
\affiliation{School of Physical Sciences, National Institute of Science Education and Research, An
OCC of Homi Bhabha National Institute, Jatni-752050, Odisha, India}
\author{Pratap K. Sahoo}
\email{pratap.sahoo@niser.ac.in}
\affiliation{School of Physical Sciences, National Institute of Science Education and Research, An
OCC of Homi Bhabha National Institute, Jatni-752050, Odisha, India}
\alsoaffiliation{Center for Interdisciplinary Sciences (CIS), NISER Bhubaneswar, Jatni-752050, Odisha, India.}
\author{Kartik Senapati}
\email{kartik@niser.ac.in}
\affiliation{School of Physical Sciences, National Institute of Science Education and Research, An
OCC of Homi Bhabha National Institute, Jatni-752050, Odisha, India}
\alsoaffiliation{Center for Interdisciplinary Sciences (CIS), NISER Bhubaneswar, Jatni-752050, Odisha, India.}

\title{Bismuth Phase Dependent Growth of Superconducting NiBi$_3$ Nanorods}


\begin{document}

\begin{tocentry}
\includegraphics[width=8.5cm]{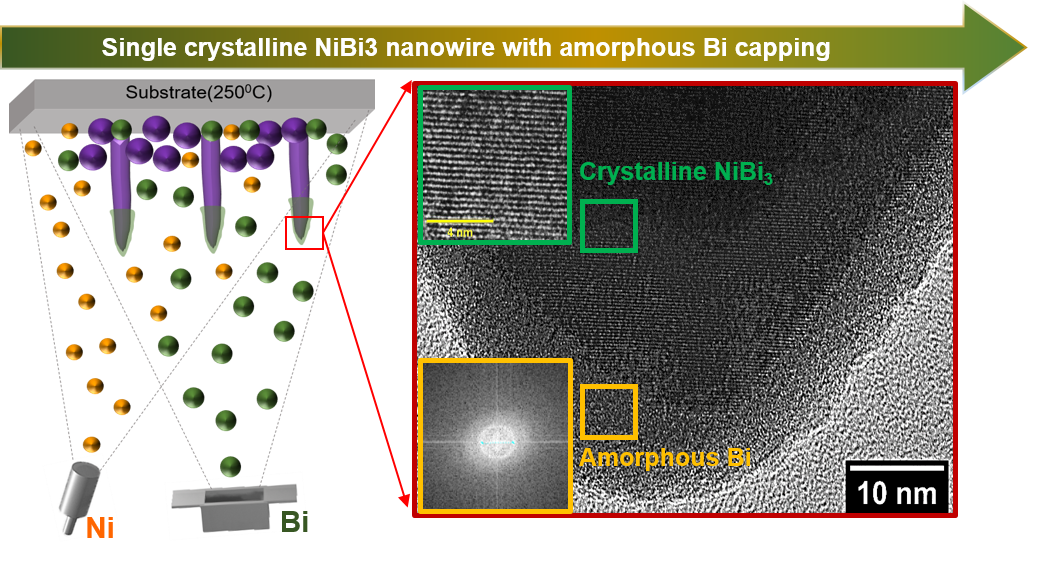}
\end{tocentry}

\begin{abstract}
  We report a study on the growth of NiBi$_3$ nanowires and nanorods during the preparation of superconducting NiBi$_3$ films by co-evaporation of Ni and Bi. We find that NiBi$_3$ films grown via co-evaporation of Ni and Bi metals achieve higher transition temperatures (4.4 K) compared even to the single crystal NiBi$_3$. However, in certain parameter space, the film surfaces were spattered with nanoscale features, such as nanowires and nanorods. Ambient temperature deposition resulted in poly-crystalline NiBi$_3$ nanorods which were controllable with the evaporation rate of Bi. Deposition at elevated temperatures promoted the emergence of long single crystalline NiBi$_3$ nanorods. High resolution transmission electron microscopy measurements confirmed the crystalline behaviour of the nanorods. We believe that NiBi$_3$ nanowires form in a process analogous to the well known vapor-liquid-solid process, as we observe an amorphous Bi cap on the nanorods. From glancing angle X-ray diffraction measurements we identify that the presence of trigonal Bi with hexagonal primitive cell in the film promotes the nucleation of nanorods. Electrical transport on a single NiBi$_3$ nanowire shows a superconducting transition of 4.3K.
  
\end{abstract}

\section{Introduction}
Numerous intermetallic compounds have been discovered possessing outstanding properties usually not found in common metals and alloys\cite{nakamura1995fundamental}. One such interesting class of intermetallic compounds, consisting of strong ferromagnetic elements such as Ni and Co, also exhibit superconductivity\textcolor{black}{\cite{liu2018superconductivity,csarli2019coexistence,tence2014cobi3,schwarz2013cobi3,wei2016first}. It is well known that ferromagnetic order strongly competes with superconducting order. Even atomically dispersed ferromagnetic impurities are known to drastically suppress the superconducting order parameter in superconductors\cite{young1999measurement}. Therefore, the existence of two macroscopic quantum orders in one single system is fundamentally a very appealing situation, apart from the technological potentials. For the same reason, the discovery of Fe based high temperature superconductors\cite{chen2008superconductivity,kamihara2008iron} sparked a lot of excitement.} In this context the bimetallic Nickel-Bismuth alloy also exhibits two stable phases, NiBi$_3$ and NiBi which are superconducting with nearly equal T$_c$ of $\sim$ 4.2 K and $\sim$ 4.1 K\cite{nedellec1985anomalous,fujimori2000superconducting,kumar2011physical,zhu2012surface,silva2013superconductivity}.  Although there have been conflicting reports on the co-existent superconductivity and magnetism\cite{fujimori2000superconducting,nedellec1985anomalous,kumar2011physical,zhu2012surface,silva2013superconductivity,gonsalves2016superconductivity,pineiro2011possible} in NiBi$_3$ system, the very fact that the superconductor is based on a strongly ferromagnetic element, is interesting in itself. Coexistence of superconductivity and ferromagnetism in Bi-Ni system was first reported by LeClair et al. in a bilayer Bi/Ni system\cite{LeClair}. For certain low thickness of Ni, using spin polarized tunnelling techniques, they observed a clear superconducting gap and a spin polarization of the conduction electrons signifying ferromagnetism. Transport measurements on polycrystalline NiBi$_3$ samples by Piñeiro et al.\cite{pineiro2011possible} and on single crystalline NiBi$_3$ samples by Herrmannsdörfer et al. \cite{herrmannsdorfer2011structure} have shown that superconducting order appears well within ferromagnetically ordered state while no Ni impurities were detected in the samples. In fact, Piñeiro et al.\cite{pineiro2011possible}have shown that the ferromagnetic ordering sets in at a much higher temperature compared to the Curie temperature of Ni. There is also theoretical evidence\cite{csarli2019coexistence} supporting the coexistence of these two orders. On the other hand, Silva et al.\cite{silva2013superconductivity} have reported that ferromagnetism in the NiBi$_3$ system appears due to the inclusion of amorphous Ni impurities. Recent high resolution surface magneto-optic Kerr effect \cite{Wang} also did not observe any ferromagnetism in single crystal NiBi$_3$ samples. Electron spin resonance experiments by Zhu et al. \cite{zhu2012surface}suggested that only a surface ferromagnetic fluctuation exists in NiBi$_3$ crystals. Therefore, the coexistence of superconductivity and ferromagnetism is still an interesting debatable topic.

Several techniques have been employed for the preparation of this bimetallic superconducting NiBi$_3$ compound in order to explore its diverse properties. For example,  Sakurai et al.\cite{sakurai2000thermoelectric} prepared polycrystalline NiBi$_3$ by co-melting high purity Ni and Bi from 900$^\circ$ C to 1100$^\circ$ C and studied the thermoelectric, thermogalvanomagnetic properties of this system. A similar method of high temperature melting of Bi and Ni was used by Fujimori et al.\cite{fujimori2000superconducting} to prepare NiBi$_3$ polycrystals and needle crystals for finding the superconducting, normal state properties of this system.

In the thin film form, it has been found that NiBi$_3$ compound forms at the interface of Ni and Bi prepared at room temperature \cite{siva2015spontaneous,siva2016superconducting,bhatia2018superconductivity,siva2017interface,liu2018superconductivity,liu2020magnetic,Vaughan2020Origin,dybkov1996growth}. The NiBi$_3$ layer formed in this way, however, is confined only to the interface. Till now there are very few reports\cite{das2023reaction} on thin films of NiBi$_3$. It is known from earlier literature from several groups that the combination of Ni and Bi is highly diffusive in nature which is due to the lower formation energy of NiBi$_3$ compared to individual Ni and Bi. As a result, the relative abundance of these two elements in a physical vapour deposition process and the available thermal energy at the substrate can drastically alter the microstructural evolution of the NiBi$_3$ films formed by physical vapour deposition process. While the strongly reactive nature of the Ni and Bi elements favors the formation of NiBi$_3$, it also leads to diverse surface texture and topographical features on the thin films in certain parameter windows, which the co-evaporation process allows to tune. Therefore, the primary motivation of this manuscript is to demonstrate this strong dependence of microstructural growth on the above mentioned parameters. We find that a higher evaporation rate of Bi, which leads to an excess of Bi impurity in the NiBi$_3$ films, triggers polycrystalline nanowire like outgrowth on the film surface. A similar effect was observed when the co-deposition of Ni and Bi was carried out at elevated surface temperatures. In this case, however, very long crystalline nanowires of NiBi$_3$ were found on the surface of the NiBi$_3$ films. The growth of these crystalline nanowires was discussed in the light of the well known vapour-liquid-solid (VLS) method. We have also identified that the presence of Bi with hexagonal primitive cell promotes such anisotropic vertical growth of NiBi$_3$ in the form of polycrystalline and single crystalline nanowires.   

\section{Experimental}
Bismuth ingots (99.99$\%$ pure) and Nickel wires (99.999$\%$ pure) were simultaneously evaporated from thermal and e-beam ports respectively in a deposition chamber maintained at a base pressure of 1$\times$10$^{-7}$ mbar. A series of thin films were prepared by varying parameters such as Bi deposition rate and substrate temperature in order to understand the effect of these parameters on the microstructure of NiBi$_3$. For the first batch of samples, the Ni deposition rate was fixed at 0.1 \AA/s while the Bi deposition rate was 0.2 \AA/s (named Sample A), 0.4 \AA/s (named Sample B), and 0.8 \AA/s (named Sample C). Another batch of co-evaporated samples was prepared by varying the substrate temperature as 150$^{\circ}$ C (named Sample D) and 250$^{\circ}$ C (named sample E) for a fixed deposition rate of Bismuth as 0.4\AA/s. \\
A field emission scanning electron microscope  (FESEM) was used to examine the surface morphology. The influence of evaporation rate of Bi and the effect of elevated substrate temperature on the phase of NiBi$_3$ was studied using a grazing incidence X-ray diffraction (GIXRD) approach using a Cu K$\alpha$ source with a Rigaku automated X-ray diffractometer at a grazing angle of 1.0 degree with a step size of 0.02deg. An energy-dispersive X-ray spectrometer (EDXS) attached to the FESEM column was used to perform elemental mapping on single nanorods. High resolution transmission electron microscopy (HRTEM) Jeol F200, operating at 200 keV was used to verify the crystalline structure of single NiBi$_3$ nanorods. Low temperature resistance meausrement was performed on a single NiBi$_3$ nanowire using cryogenic physical property measurement system (PPMS) with a minimum achievable temperature of 1.8 K.

\section{Results and discussion}

\subsection{Effect of the evaporation rate of Bismuth}
For samples A, B, and C, after attaining base vacuum in the deposition system, the evaporation rate of Ni was fixed at 0.1 \AA/s in presence of a mechanical shutter in front of the substrates. Subsequently, the evaporation rate of Bi was set, prior to opening the shutter. With the increasing rate of Bi evaporation, the Ni to Bi ratio in the vapor flux decreases, leading to excess Bi in the films. Figures 1(a), (b), and (c) display the FESEM images of samples A, B, and C, respectively, captured at the same magnification and the same tilt angle of 54 degrees from the surface of the film. 
It is observed that the surface roughness increases slightly with an increase in the evaporation rate of bismuth which is consistent with the previously reported literature\cite{mtshali2018effect,liu2021spontaneous}.
We notice a considerable number of composite nanorod-like outgrowth with a typical height between 1 to 2 $\mu$m in sample-C as shown in Fig. 1(c). Inset cross-sectional SEM image in Fig. 1(a) gives information on the thickness of the corresponding film.
 \begin{figure}
  \includegraphics[width=1\textwidth]{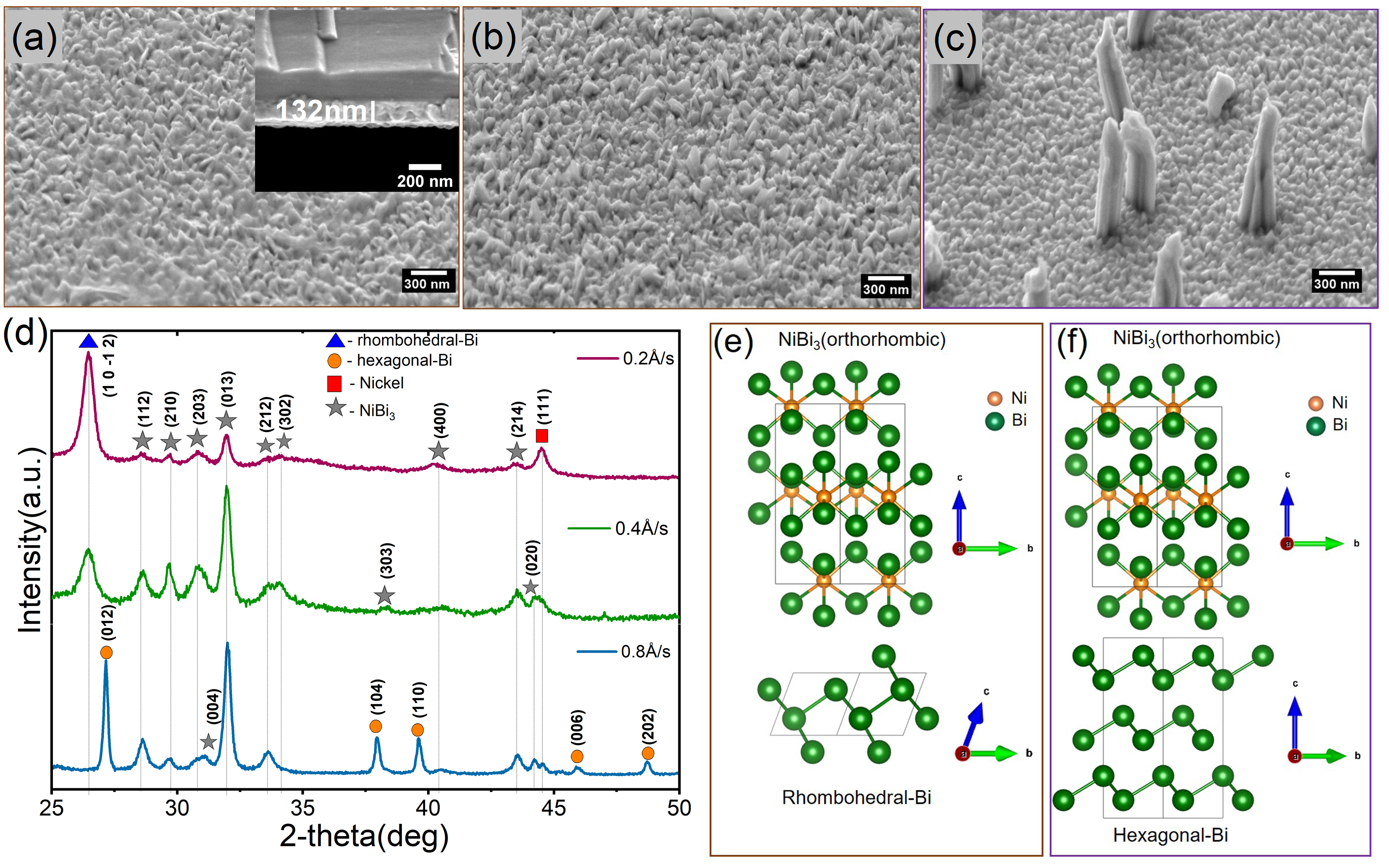}
  \caption{FESEM images of the co-evaporated samples deposited at a different evaporation rate of Bi (a) 0.2 \AA/s.Inset: cross-sectional FESEM image showing the thickness of the film, (b) 0.4 \AA/s, (c) 0.8 \AA/s, (d) GIXRD plots for all three respective samples showing polycrystalline NiBi$_3$ planes with two different phases of bismuth (e) 2D atomic representation of NiBi$_3$ orthorhombic system (a=8.884 \AA, b=4.15882 \AA, c=11.485 \AA  and $\alpha$=$\beta$=$\gamma$=90$^\circ$) over the rhombohedral-Bi where lattice parameters are a=b=c=4.798 \AA, $\alpha$=$\beta$=$\gamma$=57.423$^\circ$. (f) 2D atomic representation of NiBi$_3$ orthorhombic system over the hexagonal-Bi where lattice parameters are a=b=4.546 \AA, c=11.862 \AA and $\alpha$=$\beta$=90$^\circ$, $\gamma$=120$^\circ$.}
  \end{figure}

The nanorod like vertical outgrowth may be reminiscent of the well known vapor-liquid-solid mechanism of nanowire growth in chemical vapour deposition\cite{li2011growth,iacopi2007plasma}. Since Bismuth has a very low melting temperature of $\sim$270 $^o$C, the Bi vapour approaching the surface of the substrate is likely to be very mobile. Therefore, Bi on the surface may mimick the role of the liquid phase of the vapour-liquid-solid process to facilitate the growth of the nanorods. 

The GIXRD plot for all three samples is shown in Fig. 1(d). Multiple peaks of orthorhombic NiBi$_3$ system could be identified from these plots out of which (013) peak at 31.95$^\circ$ is the strongest in all the cases.  However, we see an interesting change in the position of the Bi peak in the sample-C compared to the sample-A and sample-B. Bi belongs to the trigonal lattice system with space group R-3m(166). The trigonal lattice system with three-fold rotational symmetry has two sub groups with (i) Rhombohedral and (ii) Hexagonal primitive cells. From Rietveld refinement, the Bi peaks appearing in sample-A and sample-B were identified as the peaks corresponding to the rhombohedral primitive cell system, hereafter termed as "rhombohedral-Bi". The lattice parameters of the rhombohedral-Bi are a=b=c=4.798 \AA, $\alpha$=$\beta$=$\gamma$=57.423$^\circ$ as per the ICSD$\#$:064703. However, in case of sample-C, the Bi peaks were identified as peaks corresponding to the hexagonal primitive cell of Bi, hereafter termed as "hexagonal-Bi". The corresponding lattice parameters are a=b=4.546 \AA, c=11.862 \AA and $\alpha$=$\beta$=90$^\circ$, $\gamma$=120$^\circ$ (JCPDS file no:85-1329). Therefore, it appears that the hexagonal axis symmetery of trigonal Bi phase may be promoting the growth of NiBi$_3$ nanorods, as these nanorods are seen only in sample-C. 

We note that the orthorhombic NiBi$_3$ has lattice parameters a=8.884 \AA, b=4.15882 \AA, c=11.485 \AA (JCPDS file no:65-0088). Fig. 1(e) and (f) show the 2D representation of the unit cell stacking of orthorhombic NiBi$_3$ as compared to the unit cell stacking of hexagonal-Bi and rhombohedral-Bi phases, as seen from the (100) zone axis. This 2D-representations modeling is done using VESTA program\cite{momma2011vesta}. The illustrations in Fig. 1(e) and (f) clearly indicates that the lattice parameters of the hexagonal-Bi are closer to the lattice parameters of orthorhombic NiBi$_3$ compared to the rhmbohedral-Bi. Therefore, the hexagonal-Bi is more likely to promote preferential growth of NiBi$_3$ nanorods. This may be the major reason for the growth of nanorods in sample-C, where hexagonal-Bi was found. The GIXRD figure also shows that when the evaporation rate of Bi increases from 0.2 \AA/s to 0.4 \AA/s, the strength of all NiBi$_3$ peaks increases, while simultaneously, the intensity of Bi peak decreases. This increase in relative intensity of NiBi$_3$ indicates that more Bi is being used in the diffusion process to form NiBi$_3$ compound\cite{das2023reaction}. The drop in the relative intensity of Ni in sample-B is also consistent with this observation. In the case of sample-C, there are other NiBi$_3$ planes such as (1 1 2), (0 0 4), (2 1 2), (2 1 4), (0 2 0) become prominent and corresponding FWHM also decreases compared to other two samples. In addition to this, another notable finding from this GIXRD plot is the NiBi$_3$ (0 2 0) plane close to Ni (1 1 1) starts to appear in sample-B giving rise to the broadening of the peak. The relative intensity of NiBi$_3$ (0 2 0) increases with further increase in the evaporation rate of bismuth to 0.8\AA/s. 

\begin{figure}
  \includegraphics[width=0.8\textwidth]{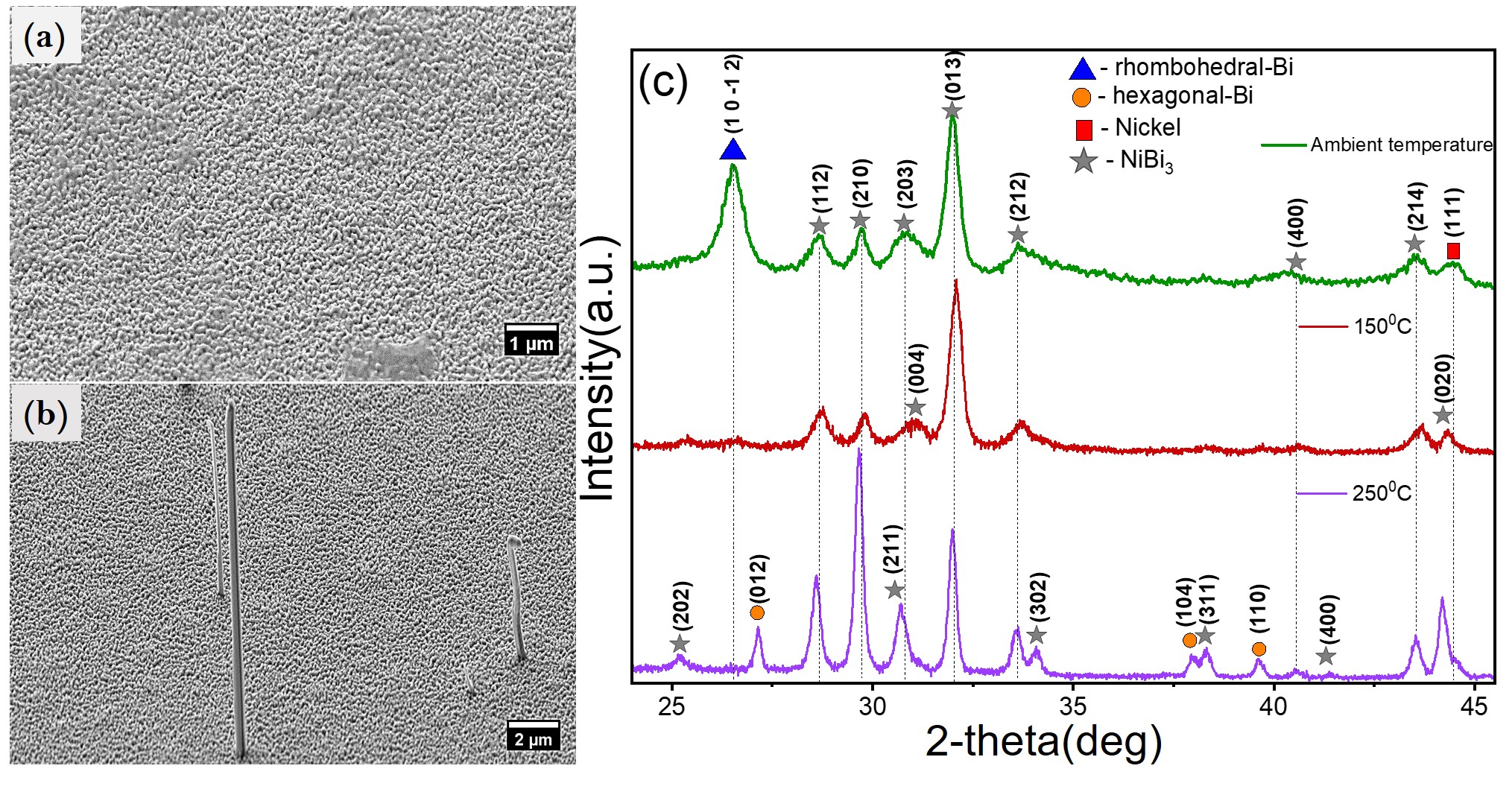}
  \caption{FESEM images of the sample deposited at different substrate temperatures (a) 150$^{\circ}$C (b) 250$^{\circ}$C. With increasing temperature, outgrown rods are found to appear significantly although in less density but higher in length than the earlier shown in Fig. 1(c). (c) GIXRD plot for the 2 respective samples compared with the as-deposited sample (top curve). The hexagonal-Bi phase is present in the sample deposited at 250$^{\circ}$C.}
\end{figure}
\subsection{Effect of substrate temperature}
In order to further ascertain the correlation between the Bi-phase and the nanowire/nanorod growth, we have prepared another set of co-evaporated samples by varying substrate temperature to 150$^{\circ}$C (sample-D), 250$^{\circ}$C (sample-E). Fig. 2(a) and (b) represent the FESEM images of the two samples deposited at elevated substrate temperatures. Fig. 2(a) shows that although the sample surface has a significant degree of surface roughness, there are no signatures of nanorods. However, very long nanowires can be seen on the surface of sample-E, in Fig. 2(b). These nanowires appear to be quite distinct from the nanorods shown and discussed in the previous section. These nanowires are much longer and less dense than the nanorods shown in Fig. 1(c). The length of nanowires varies from 10 to 25 $\mu$m with an average width of 300 nm. There are no previous reports on such longer NiBi$_3$ nanowires prepared by the physical vapour deposition process. The effect of substrate temperature has been found to be significant for the formation of nanowires\cite{mtshali2018effect}. Since Bi has a melting point of 271.4$^{\circ}$C, a sample deposited around this temperature is expected to have enough thermal energy for the crystallization of NiBi$_3$ nanowires.

Fig. 2(c) shows the GIXRD plots of samples D and E, along with the ambient temperature grown film at the same evaporation rates of Ni and Bi. In all three graphs, there are appreciable numbers of NiBi$_3$ peaks, as well as a few bismuth peaks. There is a broad peak visible for Ni (1 1 1) plane in the ambient temperature grown sample whereas, in the other two samples, the relative intensity of nickel has decreased significantly. At the same time, the intensity of the NiBi$_3$ (0 2 0) plane at 44.131$^\circ$ close to Ni (111) peak 44.45$^\circ$ has enhanced. The other noteworthy feature of the GIXRD pattern of sample-E, grown at temperatures close to the melting point of Bi, is the appearance of significantly more number of NiBi$_3$ peaks. The thin film undergoes re-crystallization process by the temperature during deposition, which has been investigated by many researchers\cite{aousgi2015effect,balakrishnan2013effect,derby2018effects,jungyoon2003effects}.
It is also known that with increasing substrate temperature, the average crystallite size increases\cite{aousgi2015effect}. Consequently, it is noticed that the FWHM of NiBi$_3$ peaks decrease significantly with increasing substrate temperature. 
 
In addition, there is an enhancement of NiBi$_3$ peaks with hkl planes (3 0 2), (3 1 1), and (0 2 0) corresponding to 2-theta values 34.036$^\circ$, 38.269$^\circ$, and 44.131$^\circ$ (close to the 2-theta value of Ni) which again confirms the re-crystallization process\cite{aousgi2015effect}. The GIXRD data in the Fig 2 shows that the 250$^\circ$C deposited Sample-E, which nucleated very long nanowires of NiBi$_3$, has Hexagonal-Bi (JCPDS file no:85-1329) phase. Similar to the previous section, we notice that congruent growth of nanowires appears only in sample-E, where the underlying Bi has a hexagonal primitive cells. In the subsequent section, we focus on a detailed analysis of the two types of NiBi$_3$ nanowires obtained in the two cases. 
\begin{figure}
  \includegraphics[width=1\textwidth]{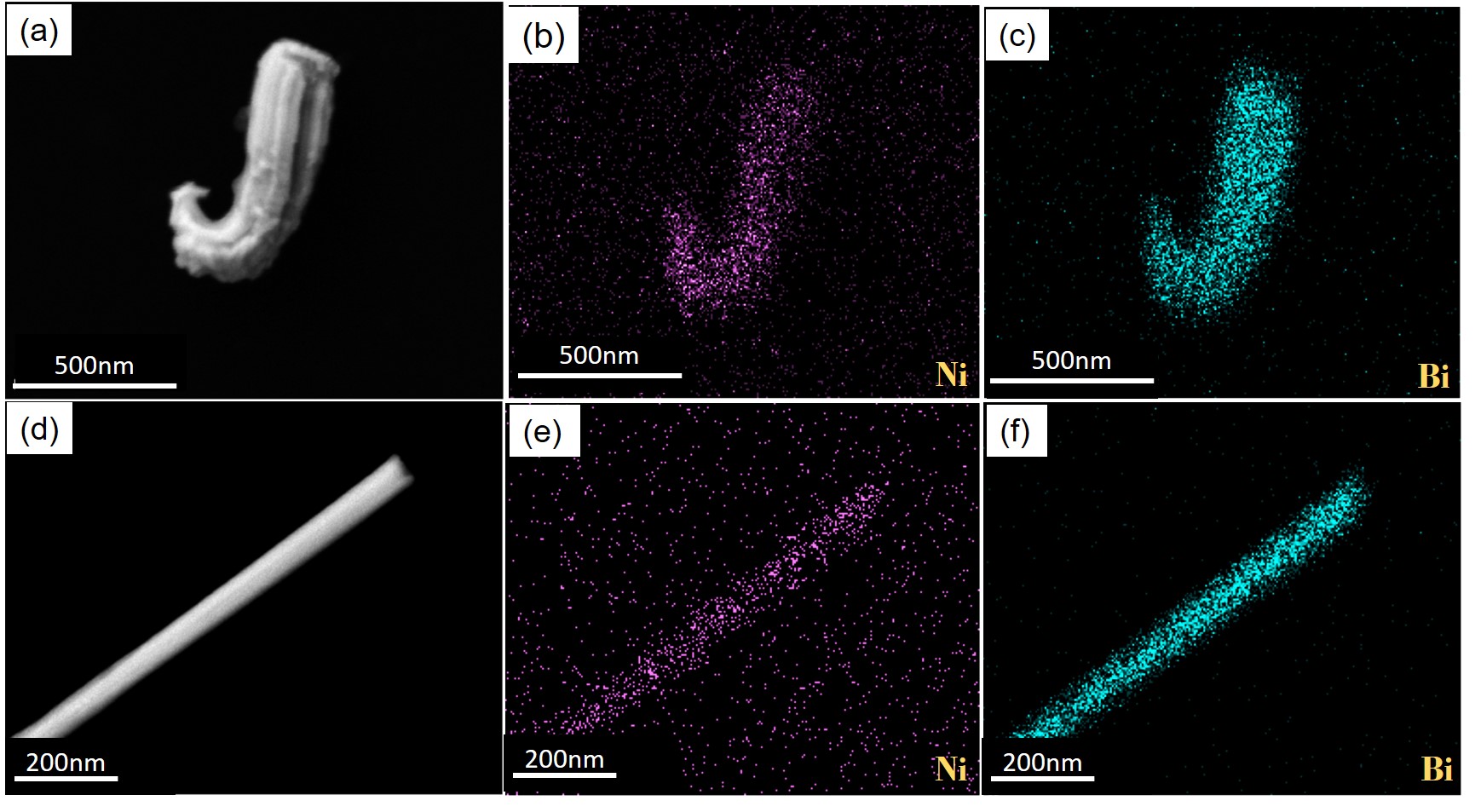}
  \caption{%FESEM image of single nanorods on the TEM grid (a)Thick and bunch kind of rod with corresponding elemental mapping of Nickel in (b) and Bismuth in (c). 
  (a) FESEM image of single nanorod on the TEM grid, with corresponding elemental mapping of (b) Ni  (c) Bi
  (d) FESEM image of single nanowire on the TEM grid, with corresponding elemental mapping of (e) Ni  (f) Bi.
%(d)Thin and long kind of rods with corresponding elemental mapping of Nickel in (e) and Bismuth in (f).
A significant amount of elemental Ni and Bi signals is found in both cases.}
\end{figure}
\subsection{Characterization of NiBi$_3$ nanowires}
\subsubsection{EDAX and HRTEM analysis}
We have performed EDAX and HRTEM on the two types of nanowires after scratching them onto a carbon-coated Cu grid to confirm the elemental distribution and crystalline nature. Fig. 3 shows the FESEM image and EDAX elemental mapping of the nanorods. Fig. 3(a) is the FESEM image for the rods that appeared on Sample-C, which was grown at ambient substrate temperatures. Fig. 3(b) and (c) depict the elemental EDAX mapping of Ni and Bi in this nanorod, demonstrating the presence of both Ni and Bi throughout. Similarly, the FESEM image of nanowires formed on Sample-E is shown in Fig. 3(d). Corresponding EDAX elemental maps are shown in Fig. 3(e) and 3(f). In this case, also the distribution of Ni and Bi is uniform throughout the nanowire. The Ni/Bi atomic percentage ratio obtained from EDAX spectra, in this case, is close to 1/3 matching well with the stoichiometry of NiBi$_3$. To understand the crystalline nature of these rods, we performed transmission electron microscopy (TEM).

\begin{figure}
  \includegraphics[width=1\textwidth]{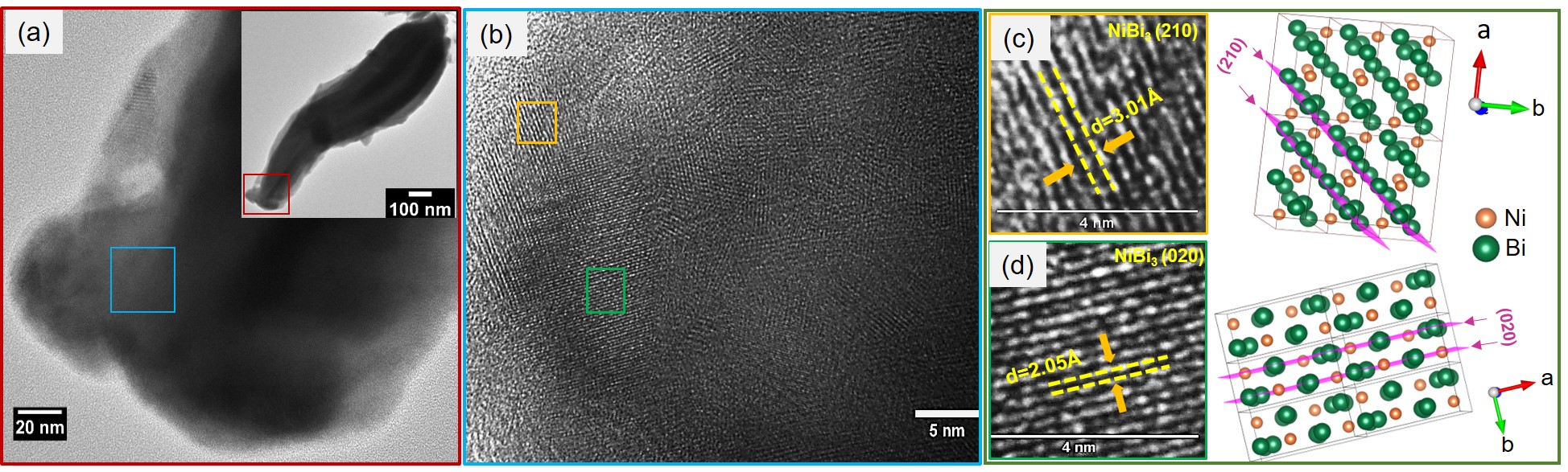}
  \caption{(a) TEM image of nanorod grown on sample-C (b) Magnified HRTEM image of the Blue marked region in (a) which shows the polycrystalline nature of this rod, (c) and (d) The magnified image of the yellow and green region in (b) showing the two planes (2 1 0)  and (0 2 0) of NiBi$_3$ with a respective 3D representation of modeling using VESTA program \cite{momma2011vesta} of these planes.}
\end{figure}

Figure 4(a) displays a TEM image of the zoomed region of the rod shown in the inset for the co-evaporated sample-C. Fig. 4(b) shows the HRTEM image of the blue square box and confirms the polycrystalline nature with different crystal orientations of the NiBi$_3$ nanorod, grown during the room temperature co-evaporation at a Bi deposition rate of 0.8 \AA/s. The most prominent crystal planes observed are (2 1 0) and (0 2 0) of NiBi$_3$ marked in the yellow and green colored box. The zoom view of the fringe pattern of the yellow area has the nearest lattice plane distance of 3.01 \AA for (210) plane of NiBi$_3$ along (001) zone axis. The corresponding schematic atomic model of the (210) lattice plane along (001) zone axis using the VESTA program\cite{momma2011vesta} is shown on the right side of Fig. 4(c). The zoomed fringe pattern of the green area corresponds to the (020) plane of NiBi$_3$, with the nearest lattice plane distance of 2.01 \AA. The schematic atomic model of the (020) lattice plane along the (001) zone axis is shown on the right side of the (020) fringe pattern. Both the lattice planes corroborate with the GIXRD peak shown in Fig. 1(d). 
\begin{figure}
  \includegraphics[width=1\textwidth]{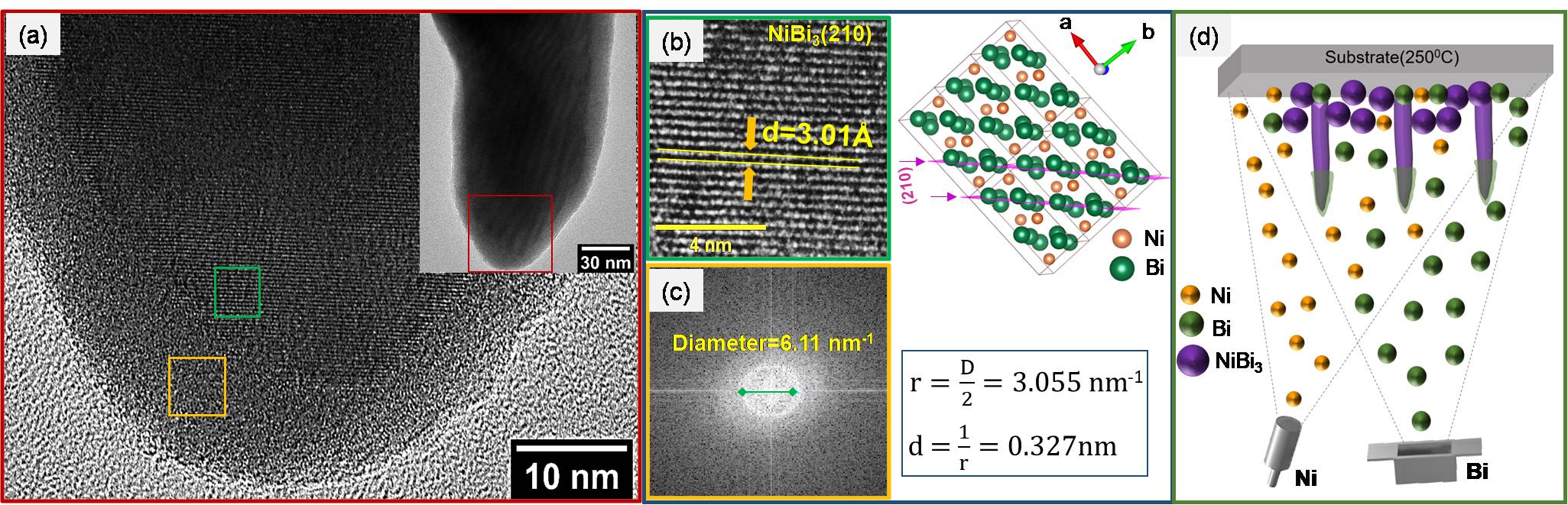}
  \caption{(a) TEM image of nanowire grown on sample-E. Inset shows the low magnification image of the nanowire (b) Magnified HRTEM image of the green marked region in(a) showing (2 1 0) plane of NiBi$_3$ with a respective 3D representation of modeling using the VESTA program. (c) The fast Fourier transform (FFT) image of the yellow marked box in (a) shows the diameter of the ring as 6.11 nm$^{-1}$. Calculated interplanar spacing as 0.327nm corresponds to Bismuth rhombohedral(012) phase. (d) Schematic of the co-evaporation process leads to nanowire growth analogous to the vapor-liquid-solid (VLS) mechanism. The cap region at the top of the nanowire is amorphous bismuth mimicking the TEM image in (a). }
\end{figure}

Similarly, Fig. 5(a) shows the HRTEM image of the nanowire formed on a co-evaporated sample deposited at an elevated temperature of 250 $^0$C (sample-E). Inset shows a part of a single nanowire of length 13 $\mu$m. We notice a clear single crystalline tip encompassed by an amorphous layer. A comparatively smaller crystalline region, marked by a green square, has been magnified in Fig. 5(b) to get precise information regarding the interplanar spacing. Here it is prominently visible that the nanowire is single crystalline. The zoomed view of the HRTEM shown in Fig. 5(b) represents the (2 1 0) planes of NiBi$_3$ along the (001) zone axis with the nearest lattice plane distance of 3.01 \AA. This NiBi$_3$ plane (210) matches with the GIXRD plot in Fig. 2(c), corresponding to the 2-theta 29.623$^\circ$. The structural atomic representation of the planes is shown by the side of the fringe pattern. Fig. 5(c) is the fast Fourier transform (FFT) of the selected area in yellow square box Fig. 5(a) which falls in the amorphous cap region. The ring-like FFT pattern shows the amorphous nature on the cap region of the nanowire. The d value, calculated from the diameter of the ring is 0.327 nm, which matches with the Bi (012) hexagonal symmetry phase. This peculiar feature at the tip of the nanowire indicates that there is a similarity in the formation of these nanowires with the well-known vapor-liquid-solid (VLS) mechanism of nanowire growth. In this mechanism, a molten metal nanoparticle works as a catalyst and promotes anisotropic growth\cite{hiruma1995growth,wagner1964vapor,wang2006mechanisms,bonu2019sub} in presence of a chemical vapor at high temperatures. Typically the molten metal catalyst remains as a cap at the top of the nanowires. The observed amorphous Bi cap, in our case, is very similar in nature\cite{su2008self,boulanger2012patterned,cui2013unusual,hatano2019germanium}. During the co-evaporation process, some Ni and Bi can interact in the vapor phase to form NiBi$_3$. In addition, some excess Bi vapor and Ni vapor also reach the surface of the substrate. Due to the low melting point of Bi, the Bi vapor reaching at the heated substrate can be extremely mobile mimicking a liquid phase. Further, Ni and NiBi$_3$ coming in contact with the apparent molten phase of Bi get catalyzed in the same manner as in VLS growth of nanowires in presence of Au droplets\cite{hannon2006influence,hijazi2020dynamics}. 
 \begin{figure}
     \centering
     \includegraphics[width=1\textwidth]{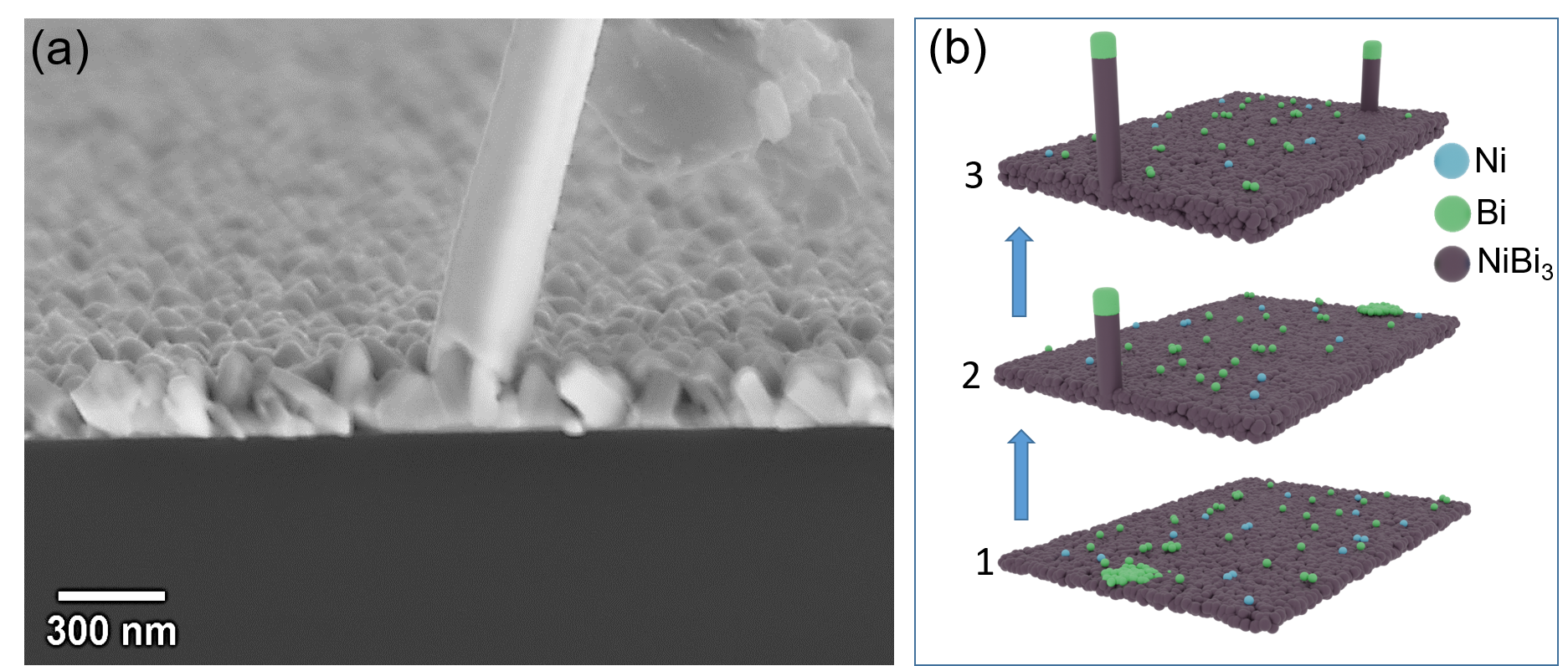}
     \caption{(a) Cross-sectional FESEM image of the sample deposited at substrate temperature of 250 $^\circ$C. A 250nm wide nanowire is shown to have grown from the base of the film. (b) A plausible model for the growth of the long nanowires is shown in this panel as snapshots at different stages of growth. In the initial stages of nanowire growth, clusters of pure Bi may appear in the film. These clusters remain in almost molten form, at substrate temperatures of 250 $^o$C, which promotes activated growth of NiBi$_3$ nanorods. }
 \end{figure}

The schematic representation of the co-evaporation process (at substrate temperature 250$^\circ$C) that results in the formation of nanowires with a Bi cap at the tip is shown in Figure 5(d). The green cap at the tip is similar to the amorphous Bi cap shown in HRTEM fig-5 (a). We want to highlight here that, in the present case, the NiBi$_3$ nanowires are much longer (10-25 $\mu$m) than the earlier literature and are single crystalline in nature when grown at 250$^\circ$C. Our observations show a clear systematism in the deposition parameters for congruent, anisotropic growth of the nanorods on the film surface. Firstly, the dependence on Bi evaporation rate (in Fig 1) shows that relatively higher concentration of Bi in the co-evaportation flux promote nucleation of NiBi$_3$ rods, albeit polycrystalline in nature. At higher evaporation rate of Bi, the chances of forming clusters/islands of pure Bi is higher. Therefore, it is very likely that the pure Bi clusters are the points of nucleation of the nanorods. This can be supported by the fact that TEM imaging shows amorphous Bi regions at the tip of the nanorods, especially well defined for the crystalline nanorods. Secondly, the temperature dependence (in Fig 2) shows that elevated temperature promotes crystalline nanorods. The average length of the crystalline nanorods was observed to be much higher than the polycrystalline rods obtained at lower temperatures. Therefore, substrate temperature strongly accelerates the anisotropic growth of rods. Since 250$^\circ$C is very close to the melting temperature of Bi, we believe that almost molten form of Bi clusters triggers the strongly anisotropic growth of NiBi$_3$ nanorods. A cross sectional FESEM image of sample-E (with very long crystalline NiBi$_3$ nanorods) is shown in Fig 6 (a). Clearly the nanorods seems to have started from the very base of the film. Based on these observations a plausible growth model of the nanowires is shown in the Fig 6(b). In the initial stages of the film growth, some clusters of Bi may appear on the surface. At growth temperatures close to the melting temperature of Bi, these clusters may act as very effective nucleation points for crystallization of NiBi$_3$ nanorods. The fact that we see an amorphous Bi cap at the tip of the nanorods says that the NiBi$_3$ crystallizes at the bottom of the molten Bi clusters formed in the initial stage of film growth. Usually, evaporation flux arriving on a substrate migrate on the surface to spread out as a thin film, However, the evaporation flux (Bi, Ni, and NiBi$_3$) arriving on almost molten Bi regions gets captured in it and helps the crystallization of NiBi$_3$ at the bottom of the cluster, very similar to the VLS mechanism widely used in the CVD growth of long semiconducting nanowires.

\subsubsection{Superconductivity in single crystalline NiBi$_3$ nanowire}
\begin{figure}
  \includegraphics[width=0.7\textwidth]{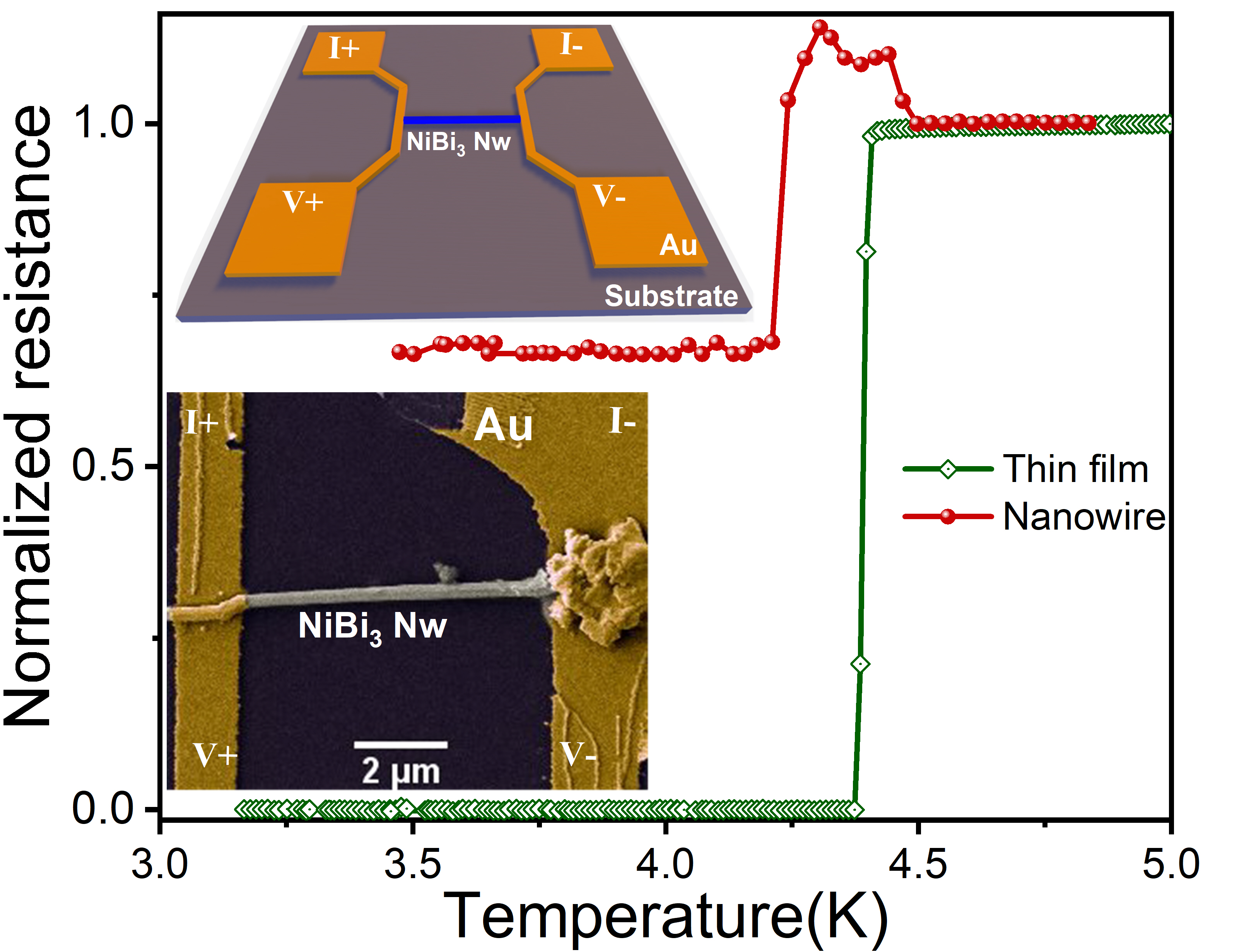}
  \caption{Resistance vs temperature measurement plot for single crystalline nanowire compared with the thin film. Inset: Schematic illustration of the transport measurement geometry, and the other one is an SEM image of the NiBi$_3$ nanowire sample used for transport measurements.}
  \end{figure}
From the analysis of SEM and HRTEM data, we find that the NiBi$_3$ nanowires obtained at a substrate temperature of 250$^\circ$ C are exceptionally long and single crystalline in nature. In this section, we verify the superconducting transition temperature of a single NiBi$_3$ nanowire produced during the growth of the NiBi$_3$ film at high temperature. For this purpose, we scratched the nanowires from the film surface onto a Si wafer coated with a 300 nm thick thermal oxide layer. Gold contact pads of 100 nm thickness were lithographically patterned at the two ends of a suitably placed NiBi$_3$ nanowire (width $\sim$ 350 nm, length=12 $\mu$m). A schematic diagram of the arrangement is shown in the inset of Fig-7. A false-color FESEM image of the actual nanowire with contact pads is also shown in the inset of the same figure. This arrangement allowed direct four-probe resistance measurement of the nanowire down to a temperature of 2 K. In the main panel of Fig-7, we compare the low-temperature resistances, normalized with the resistance at 4.75 K, of the film and the 12 $\mu$m long nanowire obtained from the same film. A bias current of 10 $\mu$A was used for both measurements. The thin film shows a clear and sharp superconducting transition at 4.4 K with a transition width of 0.05 K. The nanowire also shows a sharp transition at 4.3 K with a transition width of 0.1 K, although the resistance below the transition did not attain a zero value. This resistive nature of the nanowire is, however, a very well-known effect arising from the phase slips phenomenon in superconducting nanowires\cite{mooij2006superconducting, lau2001quantum}. Phase slip is a process by which the phase of the superconducting order parameter locally undergoes a sudden change by multiples of 2$\pi$, which momentarily creates a normal region in the nanowire. Therefore, the average effect of multiple random phase-slip events appears as a resistance below the superconducting state, at any bias current\cite{mooij2006superconducting, lau2001quantum}. Therefore, the nanowires obtained during the film growth process have been verified as NiBi$_3$ nanowires from both structural and from transport measurements.

\section{Conclusion}
In this article, we have discusses the growth of NiBi$_3$ nanorods and nanowires during the co-evaporation of NiBi$_3$ thin films. We find that the rate of evaporation of Bi and the substrate temperature, both significantly affect the growth of NiBi$_3$ nanorods and nanowires. A relatively higher rate of evaporation of Bi and a relatively higher substrate temperature seemed favorable for the formation of the nanorods. From the glancing angle X-ray diffraction measurements, we found that the presence of the hexagonal-Bi is crucial for the nucleation of the nanorods. Although nanorods were also obtained at ambient substrate temperatures at a high rate of Bi evaporation, the nanowires obtained at a substrate temperature of 250$^o$ C, were single crystalline in nature. From HRTEM images we observed that the tip of the nanowires obtained at high substrate temperature was covered with a cap of amorphous Bi. This indicates that the formation of the nanowires is analogous to the well-known vapor-liquid-solid growth mechanism of nanowires in the CVD process. In this case, although there is no liquid state as such, the high mobility of the low melting point material Bi effectively plays the role of liquid and helps nucleate the long NiBi$_3$ nanorods. It was found that the formation of nanowire occurs at random sites of the thin film with highly populated molten Bismuth relative to other spots, as seen in the cross-sectional FESEM image of the thin film. The superconducting transition of a single NiBi$_3$ nanowire, measured by four probe method, was found to be 4.3 K, matching the transition temperature of single crystal NiBi$_3$. In this work, we have identified the exact parameter window for the physical route fabrication of long, single crystal NiBi$_3$ nanowires, which will open an opportunity for quantum transport measurements in nanowires of this interesting quantum system which is a superconductor based on a strong ferromagnetic material.

\begin{acknowledgement}
The authors would like to thank the National Institute of Science Education and Research (NISER), Department of Atomic Energy, Government of India, for funding the research work through project number RIN-4001. The authors would also like thank Mr. Ritarth Chaki for help with the 3D graphics.

\end{acknowledgement}
\bibliography{Reference}

\end{document}